\documentclass[journal]{IEEEtran}
\usepackage[pdftex]{graphicx}
\usepackage{epsfig}
\usepackage[T1]{fontenc}
\usepackage{lmodern}
\usepackage{textcomp}
\usepackage{latexsym}
\newcommand{\ee}        {{\rm e}}

\newcommand{\jj}        {{\rm j}}

\begin{document}
\title{Ultra-Wideband Radar and Wireless Human Sensing}
  \author{Takuya~Sakamoto
    \thanks{Translated from IEICE Transactions Japanese Edition, vol. J103-B, no. 11, pp. 505-514, Nov. 2020 (in Japanese). \copyright 2020 IEICE}
    \thanks{
      We thank Prof.~Toru Sato of Kyoto University for his decades-long guidance and support in our study. We also thank Prof.~Alexander G. Yarovoy of Delft University of Technology, Delft, The Netherlands, and Prof.~Victor M. Lubecke and Prof.~Olga Bori\'{c}-Lubecke of the University of Hawaii, Manoa, HI, USA for their kind help with the study. We also thank our colleagues and students at Kyoto University, University of Hyogo, Delft University of Technology, and the University of Hawaii, Manoa for their cooperation with this study. Finally, we thank Kimberly Moravec, PhD, from Edanz Group for editing a draft of this manuscript. This study was supported in part by JSPS KAKENHI 21760315, 21246065, 2524905, 15K18077, 15KK0243, 19H02155, JST PRESTO JPMJPR1873, JST COI JPMJCE1307, Kyoto University SPIRITS Program, the Research and Development Project for expansion of radio spectrum resources for more efficient use of frequency resources for the future (The Ministry of Internal Affairs and Communications, Japan).}
    \thanks{T. Sakamoto is with the Department of Electrical Engineering, Graduate School of Engineering, Kyoto University, Kyoto, Kyoto, 615-8510 Japan.}
  }    
\markboth{Sakamoto: Ultra-Wideband Radar and Wireless Human Sensing}{}

\maketitle

\begin{abstract}
This study presents the progress of our recent research regarding wireless measurements of the human body. First, we explain radar imaging algorithms for screening passengers at security checkpoints that can process data faster than existing algorithms to generate accurate and high-resolution images of human bodies. Second, we introduce signal processing techniques for accurately measuring radar micro-Doppler signals associated with human motions and activities. Finally, we present radar signal processing algorithms for measuring human respiration, heartbeat, and other types of motion. Such algorithms can be applied in various fields including healthcare, security, biometric identification, and man-machine interfaces. These fields have one thing in common: prior knowledge about the human shape, motion, and physiology are effectively exploited to improve performance by optimizing the radar systems and signal processing algorithms.
\end{abstract}

\begin{IEEEkeywords}
Radar, human sensing, imaging, micro-Doppler, physiological signals.
\end{IEEEkeywords}

\IEEEpeerreviewmaketitle

\section{Introduction}
Society 5.0, which is a strategic goal of the Cabinet Office of Japan, refers to a society that can provide the necessary goods and services to the necessary people at the necessary time and in the necessary quantity \cite{society5}. To achieve this, it is important to develop next-generation sensing technologies that can continuously measure the state and condition of people accurately and inconspicuously. For example, the Apple Watch (Apple Inc., Cupertino, CA, USA), is a pioneering example of a smart watch that has sensing functions for physiological signals such as heartbeat and acceleration \cite{applewatch}. In addition, depth cameras, such as the Kinect (Microsoft Corp., Redmond, WA, USA) \cite{realsense2} and RealSense (Intel Corp., Santa Clara, CA, USA) \cite{realsense1,realsense3} can measure the position, posture, and movements of the human body with high accuracy, and such technologies are expected to be applied in many fields. In recent years, Google LLC (Mountain View, CA, USA) has been promoting Project Soli \cite{Soli}, which is a project focused on developing human body gesture recognition technology using ultra-wideband radar. The radar developed by Soli is mounted in the Google Pixel 4 smartphone.

With the development of several human body measurement technologies and the expansion of their social applications, the importance of human body sensing is rapidly increasing. Human body measurement technology that uses contact sensors (e.g., wearable devices, optical sensors, and depth cameras) are widely applied; however, the shortcomings and concerns surrounding this technology cannot be ignored. For example, contact-type devices are not suitable for long-term use because they can be a burden and cause discomfort. Therefore, people with sensitive skin, older adults, and infants need to be cautious when using such devices. In addition, unless the optical sensor is offline, privacy concerns regarding the leakage of personal information over networks cannot be eliminated.

Human body sensing using radar is attracting attention as a technology that can mitigate these shortcomings. Because radars can be used to measure the human body without physical contact, there are no concerns in terms of the discomfort experienced during wearing a device or regarding skin allergies. Radar systems involve fewer privacy-related concerns than optical cameras. In addition, microwaves and millimeter waves easily penetrate clothes and bedding; therefore, the skin surface of the human body can be measured directly even when the person being observed is sleeping in bed or wearing clothes. Therefore, radar-based human body measurements are considered to be promising. This paper introduces some of the research that we have conducted on radar-based human sensing and imaging. This paper is a revised version of a technical research report \cite{giho}.

\section{Wireless human body imaging techniques}
\subsection{Millimeter-wave radar and body scanner}
In recent years, the importance of preventing crime and terrorism to ensure safety in public places such as airports has been widely recognized. Conventional metal detectors alone cannot detect non-metal ceramic knives or illegal drugs concealed on passengers. Therefore, the importance of radar-based human body imaging technology has increased, and millimeter-wave radar-based body scanners are already in operation at security checkpoints in airports and railway stations in Japan and overseas \cite{Chen}. ProVision 2 (L3 Technologies, New York, NY, USA) is a widely used body scanner that realizes high-resolution radar imaging using an array radar that employes ultra-wideband signals (24--30 GHz) in the quasi-millimeter wave band, and it can detect the shape of objects hidden on a person. Clothing is highly permeable at this frequency band, and the waves reflected from the skin surface and concealed objects are dominant. High spatial resolution is required to achieve three-dimensional (3D) imaging of the human body shape. Therefore, for human body imaging, it is common to use an ultra-wideband antenna array in which ultra-wideband antennas are arranged in a 2D configuration.

\subsection{Radar imaging using frequency--wavenumber migration}
In this section, a planar 2D array is assumed for simplicity, but the same argument also holds for cylindrical and other curved arrays. Furthermore, we assume a monostatic radar with the same transmitting/receiving element position. The most basic radar imaging method is diffraction stack migration, which compensates for the phase rotation due to the propagation between each antenna element and each reflection point and involves a time-consuming computation in the time domain. For real-time applications, it is common to use frequency--wavenumber (F--K) migration rather than diffraction stack migration to increase the computational speed by performing the phase compensation in the frequency and wavenumber domain \cite{Tan}. In F--K migration, a 3D Fourier transform is applied to a received signal $s(x,y,t)$ in the time domain obtained for an antenna position $(x,y)$ placed on a 2D array plane ($x$--$y$ plane), to obtain F--K signal $S_{\rm FK}(k_x,k_y,\omega)$. Here, $k_x$, $k_y$, and $\omega$ are the wave number in the $x$ direction, wave number in the $y$ direction, and angular frequency, respectively. The F--K region signal is converted to $S_{\rm K}(k_x,k_y,k_z)$ using the relational expression $\omega^2/c^2=k_x^2+k_y^2+k_z^2$ in free space. Finally, a 3D inverse Fourier transform produces a 3D object image $S_{\rm S}(x,y,z)=\mathcal{F}^{-1}\{JS_{\rm K}(k_x,k_y,k_z)\}$, and 3D imaging is realized. Here, $c$ is the speed of light, and it is necessary to consider the Jacobian $J$ that accompanies the change in variables when performing the inverse Fourier transform. Imaging can be achieved quickly using a fast Fourier transform (FFT) algorithm for the Fourier transform and the inverse Fourier transform. 

\subsection{Radar imaging with a reversible transform}
The processing time of the conventional radar imaging described in the previous section depends on the speed of the FFT, but there is clearly a limit because no Fourier transform algorithm is known to be significantly faster than the FFT. Therefore, further acceleration of radar imaging has relied on hardware improvements, and it has been assumed that further acceleration by software (signal processing) would not be easy. We have instead developed a high-speed imaging method that does not rely on FFT. It relies on an approach that is fundamentally different from the conventional radar imaging framework. These methods originated from the shape estimation algorithm based on the boundary scattering transform (BST) and extraction of directly scattered waves (SEABED) method \cite{SEABED}, which is based on a reversible transform between the target shape and the feature points of the signal. Since then, we have continued to develop a variety of algorithms by increasing the imaging accuracy using a phase compensation technique \cite{phase}, application to experimental data \cite{sugino}, extension to the 3D case \cite{SEABED3D}, and extension to bistatic radar systems \cite{bistatic}. The 3D SEABED method \cite{SEABED3D,shotai} performs imaging using the equiphase plane $Z(X,Y)$ of signal $s(X,Y,t)$ received at element position $(X,Y,0)$ of the array antenna for imaging. Using the propagation delay $\tau$, $Z$ is expressed as $Z=c\tau/2$ and is the propagation distance corresponding to the propagation delay. The function $Z(X,Y)$ can be measured directly using a radar system with a high range resolution and a wide bandwidth (e.g., an ultra-wideband radar system). If the shape of the target is expressed by $z(x,y)$, we can derive the relational expression 

\begin{equation}
\left\{
\begin{array}{lcl}
X &=& x+z\partial z/\partial x\\
Y &=& y+z\partial z/\partial y\\
Z &=& z\sqrt{1+(\partial z/\partial x)^2 + (\partial z/\partial y)^2}
\end{array}
\right.
\end{equation}
which is a transform called the 3D-BST. Its inverse transform is called the 3D-IBST and is given by 
\begin{equation}
\left\{
\begin{array}{lcl}
x &=& X-Z\partial Z/\partial X\\
y &=& Y-Z\partial Z/\partial Y\\
z &=& Z\sqrt{1-(\partial Z/\partial X)^2 - (\partial Z/\partial Y)^2}.
\label{ibst}
\end{array}
\right.
\end{equation}
The 3D-IBST corresponds to the geometric optical high-frequency approximation of electromagnetic scattering. As long as the propagation distance $Z(X,Y)$ of the radar echoes are measured, the target shape can be obtained quickly by substituting it into the right-hand side of Eq.~(\ref{ibst}). This reversible transform can be applied only when there is a clear boundary between the target and the background medium. 

\subsection{Application of radar imaging to human body imaging}
\begin{figure}[bt]
\begin{center}
\epsfig{file=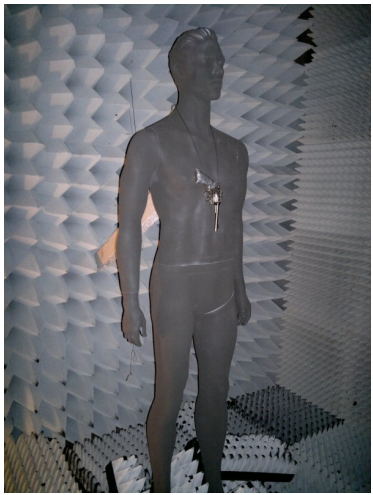,width=0.55\linewidth}
  \caption{Mannequin and handgun used in the radar imaging experiment \cite{9}.}
  \label{fig1}
\end{center}
\end{figure}

\begin{figure}[bt]
\begin{center}
\includegraphics[width=0.75\linewidth]{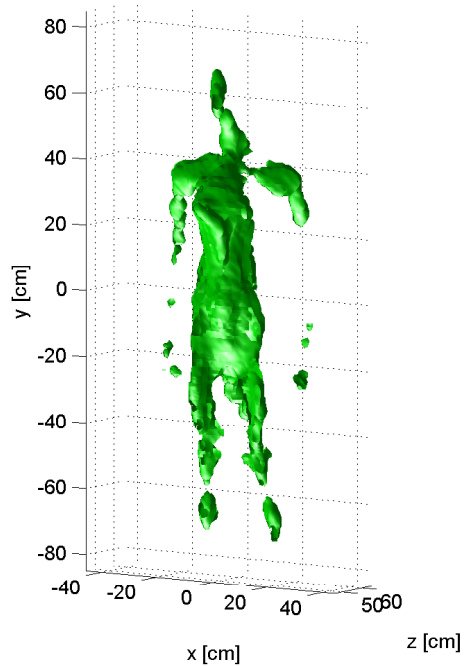}
\end{center}
  \caption{Radar image generated using the proposed RRPM algorithm \cite{9}.}
  \label{fig2}
\end{figure}

The algorithms based on these reversible transforms assume a one-to-one correspondence between the signal features and target shape. This assumption is not valid when the received signals are affected by the interference of multiple waves, leading to a deterioration in performance. This is because, when multiple waves interfere, multiple reflection points correspond to the same point in the signal space. Although increasing the signal bandwidth and the range resolution can reduce the effects of interference, the problem itself is not resolved fundamentally. With the range point migration (RPM) method \cite{RPM} and revised RPM (RRPM) method \cite{9}, the gradients calculated from multiple signal feature points are weighted and averaged to obtain the approximated derivatives that are necessary to apply the reversible transforms. Then, the estimated derivatives are substituted into the 3D-IBST in Eq.~(\ref{ibst}) to realize robust and accurate imaging even for complex shapes such as the human body. Figure \ref{fig1} shows the human body model used in a radar imaging experiment for evaluating the performance of the RRPM method, and Fig.~\ref{fig2} shows the radar image obtained using the proposed RRPM method \cite{9}. It can be seen that highly accurate estimation has been achieved, including the estimation of the handgun on the chest. The proposed RRPM was also demonstrated to achieve processing that is faster than that of the conventional F--K migration. We combined three methods, which are F--K migration, Kirchhoff migration, and 3D-IBST, to develop a final imaging algorithm that offers both high speed and high accuracy \cite{3}.

\begin{figure}[bt]
\begin{center}
\includegraphics[width=0.85\linewidth]{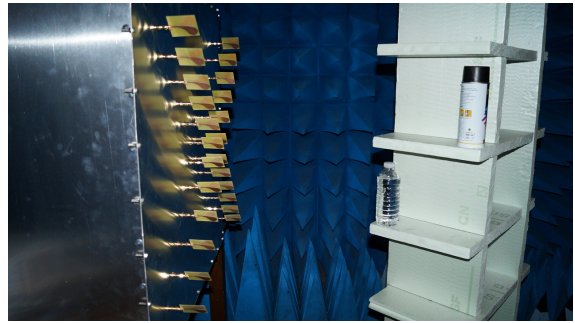}
\end{center}
  \caption{Multi-mode cavity with array antennas for simultaneous transmission and reception. The spray can and water bottle were used as targets \cite{8}.}
  \label{figThomas}
\end{figure}

\begin{figure}[bt]
\begin{center}
\includegraphics[width=0.85\linewidth]{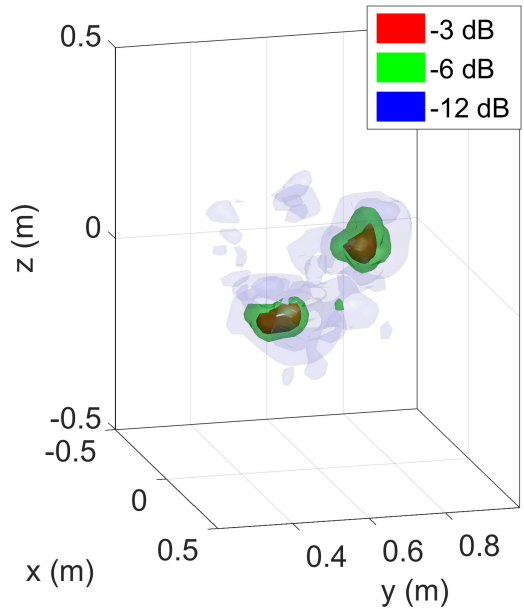}
\end{center}
  \caption{Radar image obtained using the multi-mode cavity, where the spray can and water bottle were used as targets \cite{8}.}
  \label{figThomas2}
\end{figure}

So far, we have discussed only the computational speed of radar imaging, but the speed of radar measurement itself is also important to achieve fast radar imaging. For the purpose of speeding up the measurement, \cite{8} used the fact that the time domain response of a multi-mode cavity with multi-input multi-output (MIMO) ports is uncorrelated for each pair of ports and developed a radar imaging system that realizes measurement using a 2D antenna array without mechanical scanning or switching. Figure \ref{figThomas} shows the antenna array of the measurement system. Measuring the response waveform in the cavity in advance makes it possible to acquire the measurement data of all channels at once by correlating the received signal with the response waveform of the corresponding port pair. This enabled the fast acquisition of the radar images shown in Fig.~\ref{figThomas2}.

\section{Human micro-Doppler measurements}
\subsection{Radar and micro-Doppler}
In radar measurements, the Doppler shift of the received signal has been used to measure the line-of-sight velocity of a rigid body target such as an aircraft, a ship, or a vehicle. In contrast, when measuring an object with multiple moving parts such as a drone or the human body, in addition to the representative velocity of the entire target, information about the Doppler shift caused by the movement of each part is important. The frequency shift of each part is called the micro-Doppler, and this information is known to be highly important for the identification and detection of objects. In particular, the human body has a structure wherein the torso, limbs, and head are connected, and each of these parts, although restricted in a certain way, presents highly complicated movements. Therefore, by measuring the micro-Doppler of each part of the human body, it is possible to obtain detailed information about movements that cannot be obtained by other sensors such as optical cameras, and this is an extremely powerful tool for realizing movement monitoring of the human body.

In general, to measure an accurate Doppler shift, a radar system with a narrow bandwidth is often used because of its high phase stability. In contrast, when monitoring human activity, in addition to the micro-Doppler of the torso and limbs, the position and posture of the person are also important. Therefore, ultra-wideband signals with a high range resolution are preferred in many human body measurement applications. When using an ultra-wideband radar system, the echoes from a human body can be expressed as a function of three variables: time, frequency, and range. 

\subsection{Ultra-wideband radar and texture Doppler analysis}
\begin{figure}[bt]
\begin{center}
\epsfig{file=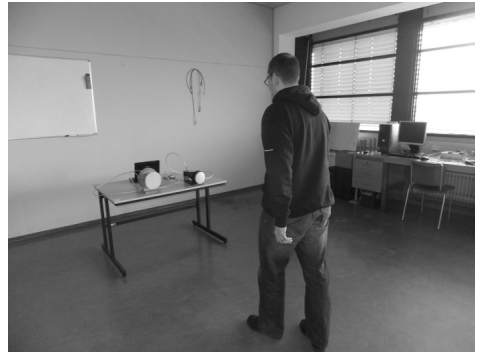,width=0.7\linewidth}
  \caption{Measurement of a walking person using an ultra-wideband radar system \cite{10}.}
  \label{fig3}
\end{center}
\end{figure}

\begin{figure}[bt]
\begin{center}
\epsfig{file=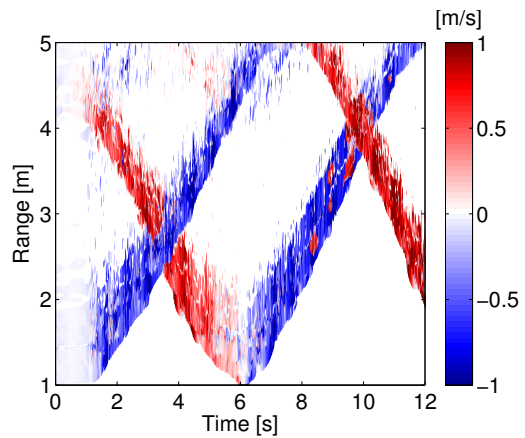,width=0.85\linewidth}
  \caption{Echoes from two walking people are displayed with their times, ranges and Doppler velocities \cite{10}.}
  \label{fig4}
\end{center}
\end{figure}

We proposed a new tool for processing an ultra-wideband radar signal in a 3D domain with the dimensions of time, frequency, and range, that is suitable for monitoring human motion \cite{5}. This approach uses prior information about the human body and motion, such as the cyclical changes in the speeds of limbs that produce the micro-Doppler. To estimate the velocity of the human motion accurately, we developed an algorithm called the texture method, which uses the texture pattern of the radar time-range image caused by the movement of the human body \cite{1}. The texture method was then extended to derive a closed-form analytical expression for estimating multiple velocities from interfering echoes without performing time-frequency analysis. This method was demonstrated to be effective in monitoring walking people \cite{10}. Figure \ref{fig3} shows the radar experiment scenario with a walking person, and Fig.~\ref{fig4} shows the resultant image generated using the proposed method, in which the velocities of two people could be estimated correctly without the Fourier transform or time-frequency analysis. In particular, their speeds are displayed in high resolution in terms of time even when their echoes interfere.

Furthermore, based on the texture method, we developed a sub-Nyquist velocity estimation method using ultra-wideband radar. With the proposed method, we succeeded in accurately estimating the velocity of a high-speed target that does not satisfy the Nyquist condition without suffering from aliasing \cite{11}. In addition, we expanded the frequency domain interferometry method that separates multiple echoes in the frequency domain using the difference in the Doppler velocities, and realized high-resolution radar imaging in conjunction with adaptive array signal processing \cite{12}. In addition, for human body imaging, we applied the 4D ESPRIT (estimation of signal parameters via rotational invariance techniques) algorithm in the four-dimensional space ($x, y, t, \omega$) using an ultra-wideband array radar to realize high resolution radar imaging \cite{Morimoto}. We believe that these high-dimensional signal processing approaches are the key to breakthroughs in the radar measurement of human movements in future.

\section{Radar-based noncontact measurement of physiological signals}
\subsection{Skin displacement and physiological information}
Small displacements (e.g., up to a few millimeters) can be seen on the skin surface of the human body as a result of respiration and heartbeats. Physiological sensing using radar can detect the Doppler shift and phase rotation of echoes to detect these small skin displacements and obtain physiological information about the target person. Assuming an ideal case where the target person is at rest and the body has only one reflection point, the distance from the antenna to the reflection point can be expressed as $d(t)=d_0+d_{\rm T}(t)+d_{\rm R}(t)+d_{\rm H}(t)$, which is a quasi-periodic function of time, where $d_0$ is the average distance to the reflection point, $d_{\rm T}(t)$ is the unconscious body movement, and $d_{\rm R}(t)$ and $d_{\rm H}(t)$ are the body displacements due to respiration and the heartbeat, respectively. Typically, the periods of respiration and heartbeat are approximately 5 s and 1 s, respectively, but there are large individual differences. Furthermore, we note that the properties of these physiological signals significantly vary over time, even for the same individual.

Assuming such a model, the complex signal obtained with quadrature detection can be written as $s(t)=A\ee^{\jj 2k d(t)}+s_{\rm DC}$. Here, $A$ is the complex amplitude, $k$ is the wave number, and $s_{\rm DC}$ is the direct current component corresponding to the leakage of the transmitted wave and the reflected wave from a stationary object, called static clutter. The distance $d(t)$ is obtained by suppressing $s_{\rm DC}$ with a static clutter removal algorithm \cite{18} and then calculating the phase. Next, it is separated into the four components stated above, and the heartbeat component $d_{\rm H}(t)$ is obtained.

\subsection{Accurate noncontact heart rate measurement}
\begin{figure}[bt]
\begin{center}
\epsfig{file=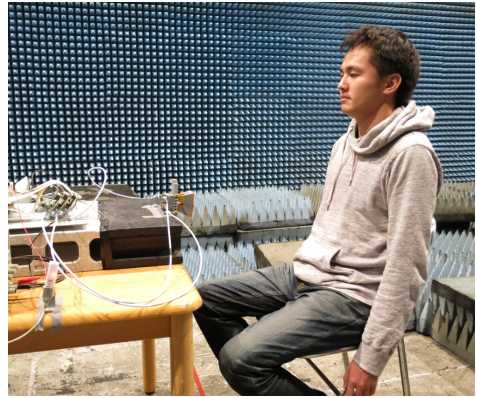,width=0.7\linewidth}
  \caption{Noncontact radar measurement of the human heartbeat \cite{7}.}
  \label{fig5}
\end{center}
\end{figure}

\begin{figure}[bt]
\begin{center}
\epsfig{file=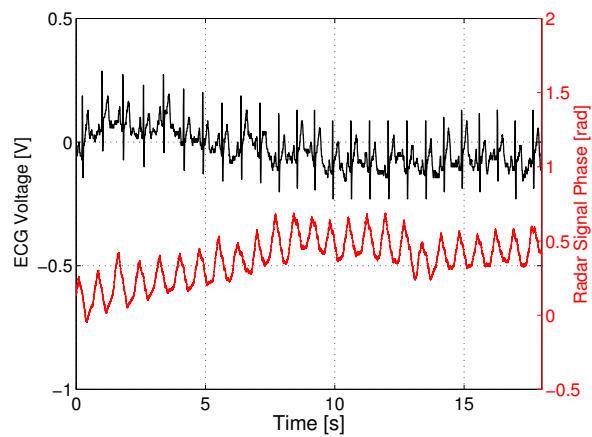,width=0.9\linewidth}
  \caption{Electrocardiogram (black) and the phase sequence of radar signal (red) of a human heartbeat \cite{7}.}
  \label{fig6}
\end{center}
\end{figure}

\begin{figure}[bt]
\begin{center}
\epsfig{file=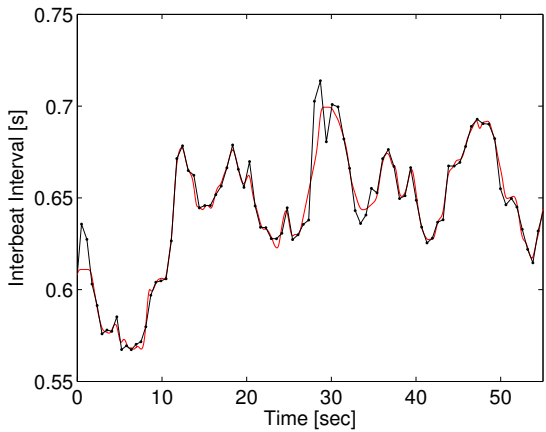,width=0.9\linewidth}
  \caption{Noncontact heart interbeat interval estimated using the topology algorithm \cite{7}. The black and red lines indicate the interbeat interval estimated using the electrocardiogram and radar, respectively.}
  \label{fig7}
\end{center}
\end{figure}

Many research groups have calculated the (average) heart rate using a Fourier transform of $d_{\rm H}(t)$, but the heart rate estimation accuracy was generally low in practice. When monitoring people, it is important to estimate the instantaneous heart rate rather than the average heart rate because the instantaneous heart rate can be converted to a mental stress index. The mechanical activity of the heart operates on the following principle: An electrical signal triggered in the sinoatrial node is transmitted to the atrioventricular node, which causes the atrium to start to contract. In contrast, there is no electrical signal that triggers relaxation (or expansion), which is slower than contraction. Thus, because the contraction and relaxation of the heart are asymmetric, it is not optimal to apply the same processing to the radar signal instances corresponding to the contraction and relaxation phases. Therefore, we focused on the asymmetry of the heart motion and developed a heart rate measurement algorithm using the topological feature points of the body displacement due to the heartbeat \cite{7}. This method is called the topology method, and its effectiveness was demonstrated in a noncontact heart rate measurement experiment using an ultra-wideband radar with a center frequency of 26.4 GHz and a 10-dB bandwidth of 726 MHz. In the experiment, it was shown that this method can measure the instantaneous heart rate with an accuracy error of approximately 1\% \cite{16}, thereby dramatically improving the practicality of wireless heart rate measurement. Figure \ref{fig5} shows the measurement scenario, where the distance from the antenna to the chest was about 60 cm. Figure \ref{fig6} shows the contact-type electrocardiogram (ECG; the red line) and the phase sequence of the radar signal (corresponding to $2kd(t)$; the black line). Using the R waves found as sharp peaks in the ECG (black line), the heart rate can be estimated accurately, whereas the radar signal (red line) does not show sharp peaks, and the instantaneous heart rate is difficult to estimate. Figure \ref{fig7} shows the instantaneous heartbeat interval sequences measured using the proposed radar-based method and ECG, where the root mean squared error of the proposed method is 7.9 ms, demonstrating the effectiveness of the proposed method. We further developed a method for efficiently suppressing the respiratory component $d_{\rm R}(t)$, and research toward practical application is progressing \cite{15}.

\begin{figure}[bt]
\begin{center}
\epsfig{file=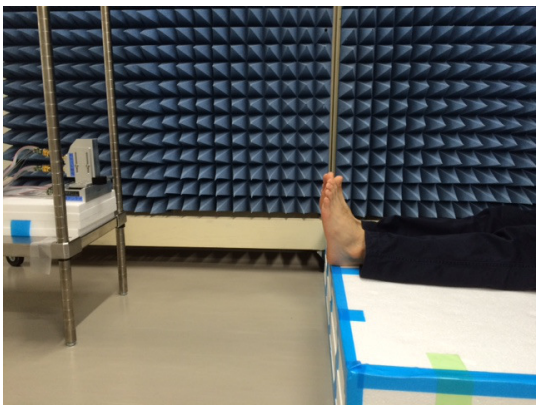,width=0.8\linewidth}
  \caption{Measurement of human soles using a millimeter-wave ultra-wideband MIMO array radar system \cite{17}. Copyright(C)IEICE 2015.}
  \label{figsoles2}
\end{center}
\end{figure}

While many research groups have focused on the measurement of the chest wall for wireless noncontact heart rate measurements, we were the first to demonstrate accurate measurement of the heart rate using the radar echo from the sole of a human foot \cite{17}. For this purpose, we introduced an ultra-wideband millimeter-wave radar with a center frequency of 60.5 GHz and a bandwidth of 1.25 GHz. Figure~\ref{figsoles2} shows the radar measurement setup targeting the sole of the participant. Subsequently, we were also the first to demonstrate noncontact heart rate measurements using the radar echo from the top of the head \cite{19}, where we used an ultra-wideband radar with a center frequency of 79.0 GHz and a bandwidth of 2.0 GHz. These radar systems have MIMO array antennas with four transmitting and four receiving elements, and we used the maximum-ratio combining technique to improve the $S/N$. This technique uses the eigenvector corresponding to the maximum eigenvalue of the correlation matrix. That the human head oscillates in synchronization with the heartbeat has been reported through the analysis of video images \cite{MIT}, but we measured this phenomenon for the first time using millimeter-wave radar. The average errors of the heart rate interval measured at the sole of the foot and the top of the head are both high (14.0 ms and 16.3 ms, respectively).

\begin{figure}[bt]
\begin{center}
\epsfig{file=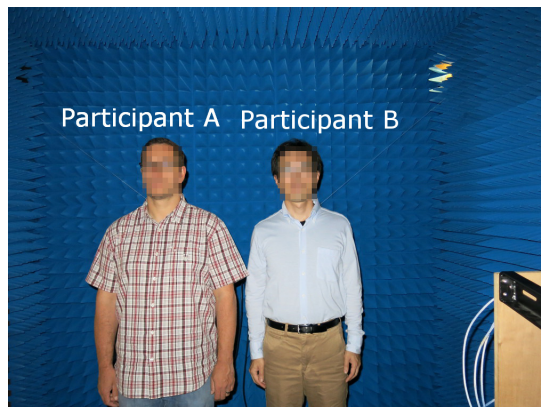,width=0.8\linewidth}
  \caption{Measurement of multiple people using an X-band array radar system \cite{13}.}
  \label{figmultiple}
\end{center}
\end{figure}

\begin{figure}[bt]
\begin{center}
\epsfig{file=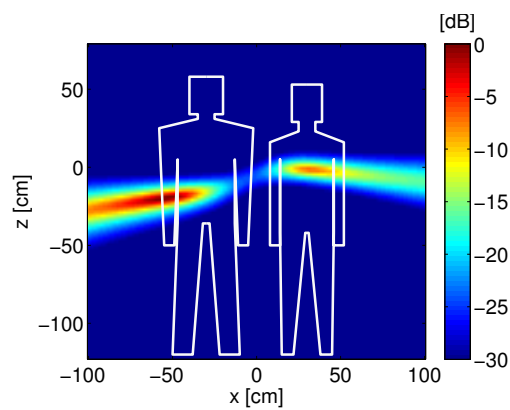,width=0.8\linewidth}
  \caption{Imaging of multiple people using an X-band array radar system \cite{13}.}
  \label{figmultiple2}
\end{center}
\end{figure}

In addition, we achieved noncontact heart rate measurement of multiple people using a four-element array antenna and adaptive array processing techniques \cite{13}. Because radar echoes that include physiological signals from multiple human bodies interfere with each other, adaptive array signal processing is used to form beams with nulls in the direction of undesired reflection points. We introduced the Capon method and the directionally constrained minimization of power method, and experimentally showed that the physiological signal of a specific individual can be extracted from interfering signals. As a result, the instantaneous heart rate of a target person can be measured accurately. Figure~\ref{figmultiple} shows the measurement setup with two standing participants, and Figure~\ref{figmultiple2} shows the obtained radar image using the Capon method.

Autonomic nervous activity can be measured using heart rate variability \cite{Pagani}--\cite{Berntson}. A time-varying heart rate is separated into low-frequency (0.04--0.15 Hz) and high-frequency (0.15--0.4 Hz) components, and the psychological state can be estimated from the intensity ratio of the two components. Although this method has been widely applied to contact-type sensors, its effectiveness has not been established for noncontact measurements. To obtain the low-frequency component, it is necessary to measure the instantaneous heart rate continuously and accurately over a time period exceeding 25 s, which is the reciprocal of the low-frequency cutoff frequency of 0.04 Hz. However, the accuracy of physiological signals estimated using a radar system depends on various factors such as the user's movement, posture, and position, which has hindered the use of radar-based noncontact measurement of autonomic nervous system activities. We proposed a technique to estimate the reliability and accuracy of the estimated instantaneous heart rate and succeeded in measuring the psychological state in terms of the mental stress index using radar signals only \cite{yamashita}. It has been experimentally shown that autonomic nervous activity can be measured accurately using this radar-based method.

\subsection{Other applications of wireless human measurement}
\begin{figure}[bt]
\begin{center}
\epsfig{file=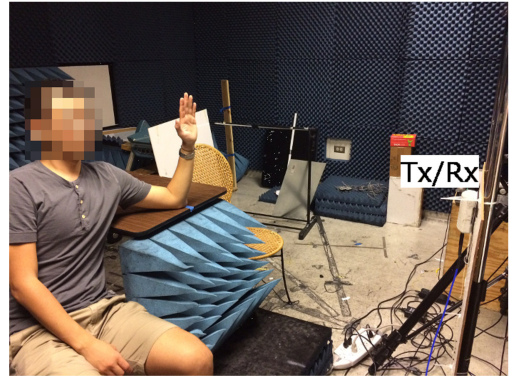,width=0.75\linewidth}
  \caption{Measurement of a person performing gestures using a 2.4-GHz Doppler radar system \cite{20}.}
  \label{fig8}
\end{center}
\end{figure}

In a smart society, it is expected that the ``Internet of Things'' technology will enable a large number of sensors to constantly monitor users and provide comfortable and convenient services. For that purpose, in addition to the measurement of physiological signals, personal identification will be important. To avoid privacy concerns, sensors used for personal identification should not involve the use of cameras or microphones, which makes noncontact sensing using radar systems promising. To achieve personal identification using a radar system alone, we measured the respiration $d_{\rm R}(t)$ of multiple seated subjects using a 2.4-GHz continuous wave radar, and the target user was accurately identified using a technique based on a neural network \cite{14}. Next, to achieve the personal identification of people in motion, we measured walking and sitting movements with a radar system and converted the received signals to micro-Doppler components that are associated with the motion patterns. Personal identification was then performed using a convolutional neural network. With this method, we succeeded in identifying all six subjects with an accuracy of 93.3\% \cite{wabunc,arxiv}. In addition, with the practical application of Google Soli mentioned in the Introduction, radar-based gesture identification has become increasingly popular. Most of these studies apply time-frequency analysis to the received signals to obtain spectrogram images for identification. However, because gesture identification technology is used for man-machine interfaces, fast computation and real-time operation are required, for which time-consuming time-frequency analysis is not preferable. Therefore, we developed a time-domain approach with a convolutional neural network applied directly to a time-domain I--Q plot and achieved the quick and accurate identification of hand gestures \cite{20}. Figure~\ref{fig8} shows the setup of the gesture identification experiment. With the proposed method, we succeeded in identifying six types of gesture with an accuracy of 91.3\% using only time-domain signals. As described above, various forms of information can be monitored remotely by measuring the movement of a human body in a noncontact manner. Applications for wireless human body sensing are not limited to those described here, and there is the potential for them to spread to all aspects of our life when linked with smartphones and other devices.

\section{Conclusion}
This study discussed various aspects of the application of radar sensing to human measurement. As for radar imaging, we focused on the boundary surface of the human body and developed algorithms based on reversible transforms. We achieved both fast processing and high spatial resolution using ultra-wideband radar systems. As for human activity monitoring using radar, we developed techniques using prior knowledge about human motion patterns, which led to novel algorithms. As for the radar monitoring of physiological signals, we developed techniques to measure the heartbeat and respiration accurately using radar, which was further extended to the monitoring of multiple people with ultra-wideband array radar systems. As exemplified in these cases, radar-based human measurement has numerous applications in fields such as healthcare, medicine, security, and man-machine interfaces, and their market size is expected to continue to grow partly owing to the trend of health consciousness and a compact radar system on a chip that can be implemented in wearable devices and smartphones. In addition, along with the increasing awareness of privacy protection, radar sensing is expected to play an important role in monitoring people without cameras or microphones. Although the research topics covered in this study have diverse applications, they are all related to the combination of electromagnetic waves and human physiology, and have achieved fast computation, high accuracy and high resolution by effectively exploiting the properties of both sensing engineering and biomedical engineering. In the future, we hope that additional studies in these topics regarding wireless human body sensing will contribute to the establishment of a new research field that integrates sensing engineering and biomedical engineering.

\begin{IEEEbiography}[{\includegraphics[width=1in,height=1.25in,clip,keepaspectratio]{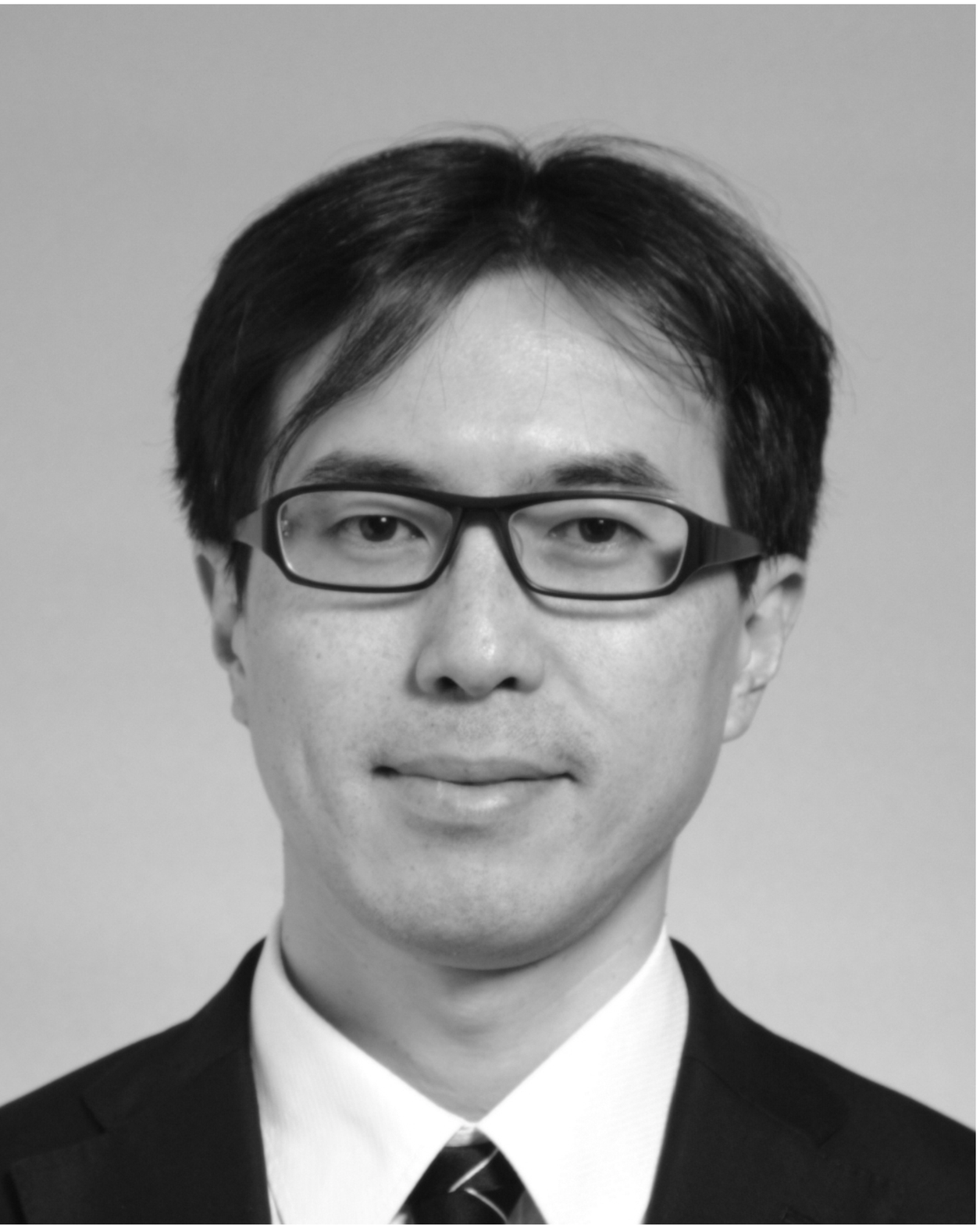}}]{Takuya Sakamoto}received a B.E.~degree in electrical and electronic engineering from Kyoto University, Kyoto, Japan, in 2000 and M.I.~and Ph.D.~degrees in communications and computer engineering from the Graduate School of Informatics, Kyoto University, in 2002 and 2005, respectively.

From 2006 to 2015, he was an Assistant Professor at the Graduate School of Informatics, Kyoto University. From 2011 to 2013, he was also a Visiting Researcher at the Delft University of Technology, Delft, The Netherlands. From 2015 to 2018, he was an Associate Professor at the Graduate School of Engineering, University of Hyogo, Himeji, Japan. In 2017, he was also a Visiting Scholar at the University of Hawaii at Manoa, Honolulu, HI, USA. From 2018, he has been a PRESTO Researcher at the Japan Science and Technology Agency, Kawaguchi, Japan. Currently, he is an Associate Professor at the Graduate School of Engineering, Kyoto University. His current research interests are system theory, inverse problems, radar signal processing, radar imaging, and the wireless sensing of vital signs.

Dr. Sakamoto was a recipient of the Best Paper Award from the International Symposium on Antennas and Propagation (ISAP) in 2012 and the Masao Horiba Award in 2016. In 2017, he was invited as a semi-plenary speaker to the European Conference on Antennas and Propagation (EuCAP) in Paris, France.
\end{IEEEbiography}
\end{document}